\documentclass{aa}  
\usepackage{amssymb}
\usepackage{graphicx}
\usepackage[authoryear]{natbib}
\usepackage[figuresright]{rotating}

\usepackage{txfonts}
\newcommand{\teff}{\ifmmode T_{\rm eff} \else T$_{\mathrm{eff}}$\fi}
\newcommand{\logg}{\ifmmode \log g \else $\log g$\fi}
\newcommand{\lL}{\ifmmode \log \frac{L}{L_{\odot}} \else $\log \frac{L}{L_{\odot}}$\fi}

\newcommand{\kms}{km s$^{-1}$}
\newcommand{\msun}{\ifmmode M_{\odot} \else M$_{\odot}$\fi}
\newcommand{\zsun}{\ifmmode Z_{\odot} \else Z$_{\odot}$\fi}
\newcommand{\lsun}{\ifmmode L_{\odot} \else L$_{\odot}$\fi}
\newcommand{\rsun}{\ifmmode R_{\odot} \else R$_{\odot}$\fi}
\newcommand{\qh}{\ifmmode Q_{\rm H} \else $Q_{\rm H}$\fi}
\newcommand{\qhei}{\ifmmode Q_{\ion{He}{i}} \else $Q_{\ion{He}{i}}$\fi}

\newcommand{\mum}{\ifmmode \mu m \else $\mu m$\fi}

\begin{document}
   \title{POLLUX: a database of synthetic stellar spectra}

   \subtitle{}

   \author{
          A. Palacios\inst{1} 
          \and
          M. Gebran\inst{2,1}
          \and
          E. Josselin\inst{1}
          \and
          F. Martins\inst{1}
          \and
          B. Plez\inst{1}
          \and
          M. Belmas\inst{1}
          \and
          A. L\`ebre\inst{1}
          }

   \offprints{A. Palacios: palacios AT graal.univ-montp2.fr}

   \institute{GRAAL--CNRS, Universit\'e Montpellier II -- UMR 5024, Place Eug\`ene Bataillon, F--34095, Montpellier Cedex 05, France \\
         \and
             Departament d'Astronomia i Meteorologia, Universitat de Barcelone, Mariti i Franqu\`es 1, E--08028, Barcelona, Spain\\
             }

   \date{}

\authorrunning{A. Palacios et al.}
\titlerunning{The POLLUX database}

 
  \abstract 
  {} 
  {Synthetic spectra are needed to determine
  fundamental stellar and wind parameters of all types of stars. They are
  also used for the construction of theoretical spectral libraries helpful
  for stellar population synthesis. Therefore, a database of theoretical
  spectra is required to allow rapid and quantitative comparisons to
  spectroscopic data. We provide such a database offering an
  unprecedented coverage of the entire Hertzsprung-Russell
  diagram.} 
  {We present the POLLUX database
  of synthetic stellar spectra. For objects with T$_{\rm eff} \le$ 
  6\,000 K, MARCS atmosphere models are computed and the program
  TURBOSPECTRUM provides the synthetic spectra. ATLAS12 models are computed
  for stars with 7\,000 K $\le T_{\rm eff} \le$ 15\,000 K. SYNSPEC gives
  the corresponding spectra. Finally, the code CMFGEN provides atmosphere
  models for the hottest stars ($T_{\rm eff} >$  25\,000 K). Their
  spectra are computed with CMF\_FLUX. Both high resolution (R$>$150\,000)
  optical spectra in the range 3\,000 to 12\,000 \AA\ and spectral energy
  distributions extending from the UV to near--IR ranges are
  presented. These spectra cover the HR diagram at solar metallicity. }  
  {We propose a wide variety of synthetic
  spectra for various types of stars  in a format that is compliant with th
  Virtual Observatory standards. A user--friendly web interface allows
  an easy selection of spectra and data retrieval. Upcoming
  developments will include an extension to a large range of metallicities
  and to the near--IR high resolution spectra, as well as a better
  coverage of the HR diagram, with the inclusion of models for Wolf-Rayet
  stars and large datasets for cool stars. The POLLUX database is
  accessible at http://pollux.graal.univ-montp2.fr/ and through the Virtual
  Observatory.}
{}


   \maketitle


\section{Presentation and aim of the database}
\label{s_intro}

POLLUX is a database of synthetic stellar spectra developed
at the GRAAL Laboratory (Universit\'e Montpellier II - CNRS). Its aim is
to provide a comprehensive library of theoretical stellar spectra with
a broad coverage of the atmospheric parameters (effective temperature
\teff, gravity \logg\ and metallicity [Fe/H]). Both high
resolution optical spectra and spectral energy distributions are
provided for O to M stars.

POLLUX spectra are expected to be useful to astrophysicists for stellar or
(extra--)galactic applications in several respects: accurate determination
of fundamental properties of stars, abundance determinations, radial
velocities and stellar dynamics, tests of the current state-of-the-art
model atmospheres, stellar population synthesis. POLLUX spectra can also be
used for teaching purposes oriented toward spectroscopy and model
atmospheres.\\ In a framework where various grids of model atmospheres,
SEDs and synthetic spectra computed with the MARCS\footnote{{\it
http://www.marcs.astro.uu.se/}}, ATLAS9\footnote{{\it
{\tiny http://www.mpa-garching.mpg.de/PUBLICATIONS/DATA/SYNTHSTELLIB/synthetic\_stellar\_spectra.html}}} and CMFGEN\footnote{{\it
http://kookaburra.phyast.pitt.edu/hillier/web/CMFGEN.htm}} codes
respectively are available on the internet, the originality of the POLLUX
database is manyfold. It is an evolutive interface that combines synthetic
spectra and spectral energy distributions based
on these three codes using the
same nomenclature, the same description and a common access
interface that is in compliance with the Virtual Observatory
standards. Among the
added values are the possibility of selective retrieval, an option that is
not commonly proposed, and that allows for the user to pre-select the data
needed, and the possibility to visualize prior to any download the spectra and their attached header files. Finally, the POLLUX database
is the only one providing CMFGEN based High Resolution Synthetic Spectra for O-type stars and soon also for Wolf-Rayet
stars.

In Sect.\ \ref{s_database} we describe the content of the
database. Then, in Sect.\ \ref{s_codes}, we present briefly the
atmosphere codes (MARCS, ATLAS12 and CMFGEN) used to produce the
synthetic spectra. Sect.\ \ref{s_web} describes the web interface to
handle and retrieve the synthetic spectra. Finally, the orientation of
the POLLUX database towards the Virtual Observatory is presented in
Sect.\ \ref{s_VO} and future developments are listed in
Sect.\ \ref{s_future}.


\section{The database content}
\label{s_database}

The database includes two types of spectra:

\begin{itemize}

\item[$\bullet$] {\it High Resolution Synthetic Spectra (HRSS data)}: they cover the optical range, 
from 3\,000 to 12\,000 \AA. Their resolution -- $R = \frac{\lambda}{\delta \lambda}$ -- is model-dependent, but always $>$150\,000. 
In practice, CMFGEN spectra have $R$=150\,000, and MARCS/ATLAS spectra have a constant wavelength step of 0.02 \AA\, 
leading to $R >$ 150\,000. POLLUX provides both absolute fluxes and spectra normalized to the 
theoretical continuum. The typical size of a HRSS ASCII file is 5.7 MB (CMFGEN), 15.4 MB (MARCS) or 14 MB (ATLAS).

\item[$\bullet$] {\it Spectral Energy Distributions (SED data)}: covering the wavelength 
range 910 -- 200\,000 \AA~ (50 -- 200\,000 \AA\ for the hottest stars), they have a resolution of 20\,000. 
The typical size of a SED ASCII file is 3.4 MB (CMFGEN), 3.2 MB (MARCS) or 1.6 MB (ATLAS).

\end{itemize}

\begin{figure*}[th]
\centering
\includegraphics[width=18cm,angle=0]{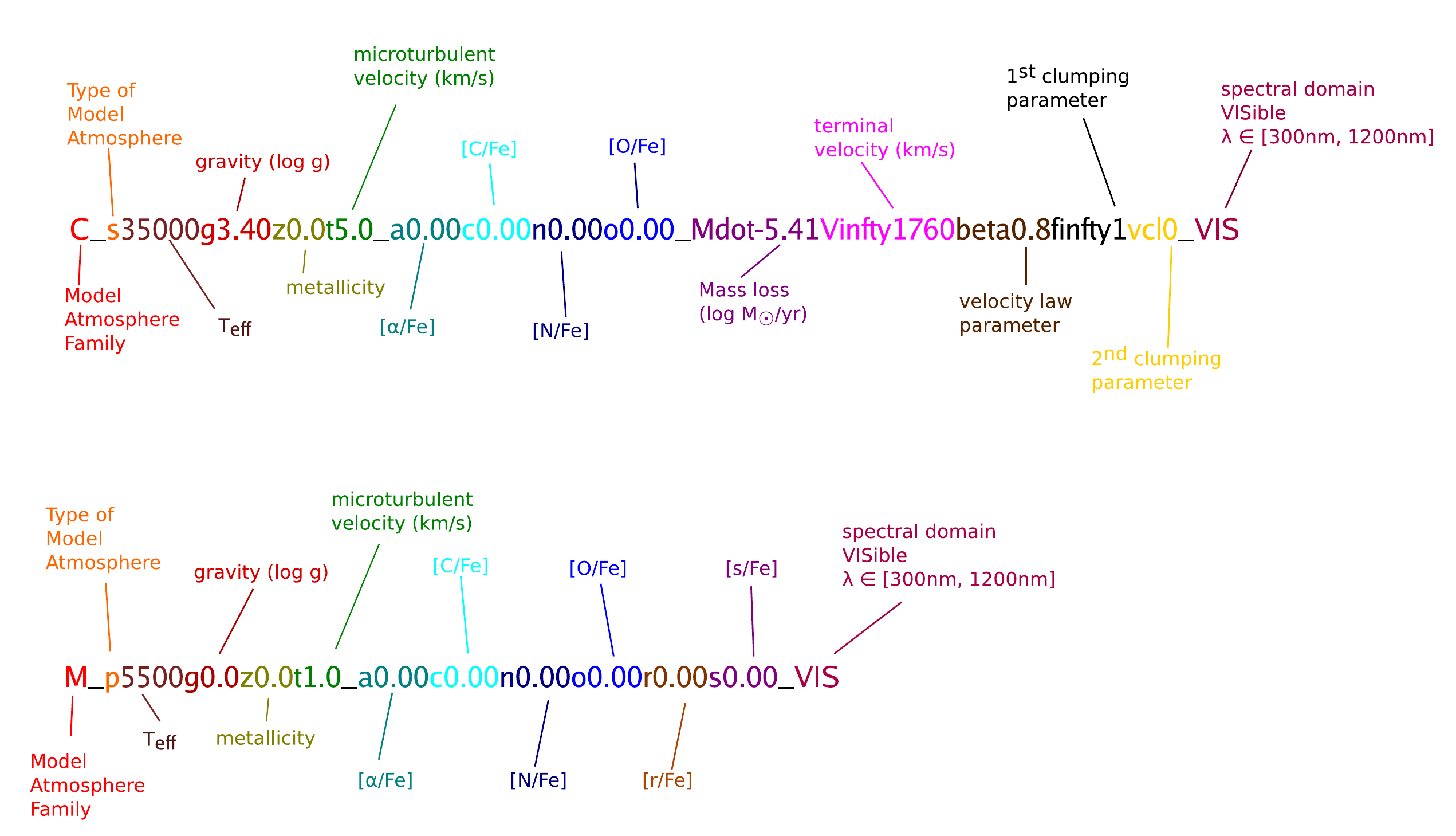}
\caption{Name nomenclature for the CMFGEN (upper line) and MARCS or ATLAS (lower
  line) files. The High Resolution Synthetic Spectra (HRSS) will be given
  the extension {\bf .spec}, the SED the extension {\bf .sed} and the
  associated headers will have {\bf .spec.txt} and {\bf .sed.txt} respectively.}\label{fig_name}
\end{figure*}

\noindent A header file is attached to each data set (HRSS or SED). It
contains a set of descriptors characterising stellar spectra (file
structure and curation\footnote{Curation includes all information
  concerning the data sets that ensures they are available for
  discovery and re-use in the future. Number version of the code, data
  producer, date of production are part of the curation information.}
information). Also included in the header is specific information on
the synthetic data (code, input physics, physical parameters
characterising the spectrum and SED).  This header file is designed to
be useful within the framework of the Virtual Observatory (see
Sect.~5).

The nomenclature adopted to label the filenames is
presented in Fig.\ref{fig_name}. It is composed of the main parameters involved in
the spectrum computation (model and data types, stellar parameters,
specific abundances) and/or of parameters necessary to access the
spectrum in the database through the POLLUX request form (see
Section~4).\\

\begin{figure}[ht]
\label{teff_logg_FeH0}
\centering
\includegraphics[width=9cm]{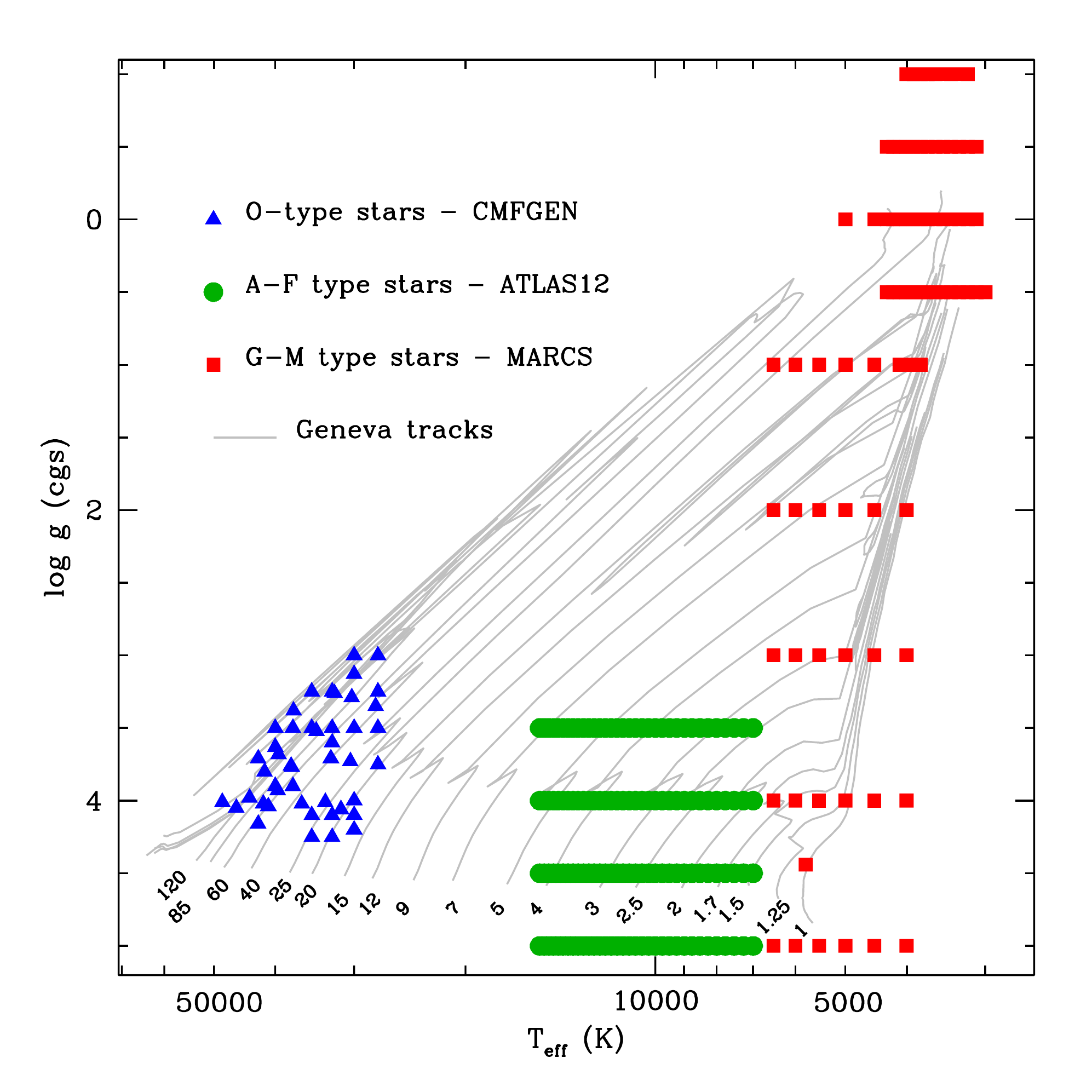}
\caption{Coverage of the (\teff - \logg) plane by the Pollux data (SED and
  HRSS) at solar metallicity (as of December 2009). Geneva stellar
  evolution tracks at Z = 0.02 with overshooting and normal mass loss from
  Schaller et al. (1992) are shown in grey. The masses, in solar mass
  units, are given on the plot.}
\end{figure}

The current (Winter 2009/2010) version of the POLLUX database contains
synthetic spectra (HRSS and SED) at solar metallicity ([Fe/H] =
0.0\footnote{The absolute solar metallicity (in mass fraction) depends on
  the adopted solar chemical composition and may vary among the data
  provided in the POLLUX database. This information is however accessible
  through the header attached to each data file.}) representative of all the parts of the HR,
with different refinement of the \logg-\teff~ plane coverage according to
the spectral type:
\begin{itemize}
\item For G to M spectral types, the HRSS and SED based on MARCS model
  atmospheres span the domains  \teff~$\in$ [3000 $K$; 6000 $K$] and \logg~
  $\in$ [-1; 5] with steps of 500 $K$ and  1.0 dex respectively. The M spectral type is present in the database through data
  representative of red supergiants (luminosity class I and M spectral
  type; see Fig.\ \ref{teff_logg_FeH0} for current coverage).
\item For late B, A and F spectral types, the HRSS and SED based on ATLAS12 model
  atmospheres are provided in the domains \teff~$\in$ [7000 $K$; 15000 $K$]
  and \logg~$\in$ [3.5; 5] with a refinement of 250 $K$ in \teff~ step and of 0.5 dex
  in \logg~ step. 
\item For O spectral type, the HRSS and SED based on CMFGEN model
  atmospheres have an irregular coverage of the (\logg,\teff) plane with
  \teff~ranging from 27500 $K$ to 49091 $K$ and \logg~ $\in$ [3.0; 4.250]
  (see Fig.\ref{teff_logg_FeH0} for current coverage). These data partly
  follow evolutionary tracks and have been computed for the analysis of
  specific real stars.
\end{itemize}

\noindent In addition to this, for late-B, A and F type stars, data based on the ATLAS12 model atmospheres are
also provided at [Fe/H] = -0.5 and [Fe/H] = -1.0 with the same coverage of
the (\logg,\teff) plane as at [Fe/H] = 0.0 (see
Fig.\ref{teff_logg_atlas}). Some additional data with non-solar [C/Fe],
[N/Fe] and [O/Fe] are also included for O stars, based on CMFGEN model
atmospheres, and for G and K stars based on MARCS model atmospheres.\\

The content will evolve in the near future, with the aim to fully cover the HR diagram
in a very wide metallicity range.

\begin{figure}[ht]
\centering
\includegraphics[width=9cm]{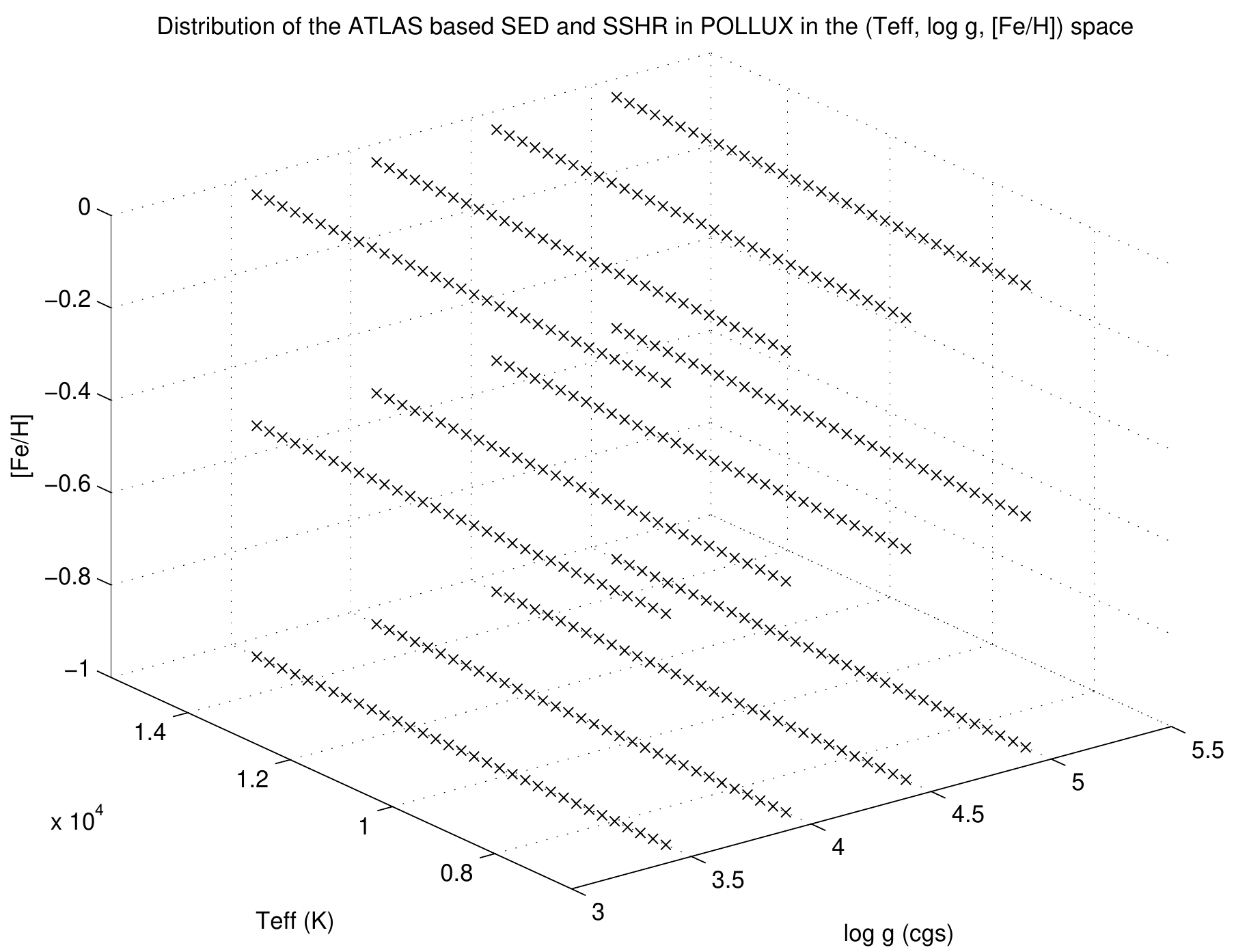} \label{teff_logg_atlas}
\caption{Coverage of the (\teff, \logg, [Fe/H]) space by the Pollux data
  (SED and HRSS) associated to ATLAS12 model atmospheres (as of December
  2009). Data are available for three different metallicities : [Fe/H] =
  0.0, -0.5 and -1.0.
}\label{teff_logg_atlas}
\end{figure}


\section{Atmosphere codes and synthetic spectra}
\label{s_codes}

Our goal is to provide the best synthetic spectra across the entire HR
diagram.  To do so, we have selected three different atmosphere codes:
MARCS, ATLAS12, and CMFGEN. These codes are best suited to model G to M, S
and C stars, late B, A and F stars, and O and Wolf-Rayet stars
respectively. To date there is no overlap in the POLLUX database between
the temperature domain covered by each code (see previous section and
Fig.~\ref{teff_logg_FeH0}). In particular there are no spectra available
for B-type stars, the modeling of the atmospheres of these objects
requiring different physics than those incorporated in the three
aforementionned codes.
Each code computes the atmospheric structure from which a formal solution of the radiation transfer is made using secondary tools: 
TURBOSPECTRUM, SYNSPEC and CMF\_FLUX. Below, we give a short description of
the assumptions and specificities of each code. We also present the scientific
  use that may be made of the spectra present in the database.

\subsection{MARCS and TURBOSPECTRUM}
\label{s_marcs}

For stars cooler than  6\,000 K, models were computed with the atmosphere
code MARCS \citep{gus75}. The most recent version is described in
\citet{gus08}. MARCS provides LTE plane--parallel or spherical models
including line--blanketing and convection. The hydrostatic structure takes
into account gas, radiation and turbulent pressure. It
assumes a constant turbulent velocity. For the models used to generate the
spectra available in POLLUX, a value of 1 km s$^{-1}$ was used for all the
models except those for red supergiants, for which the adopted value was 2 km s$^{-1}$. The total flux is the sum of the radiative
and convective fluxes. Convection is treated using the mixing length
formalism (hereafter MLT) developed by \citet{henyey65}. A value of 1.5 was chosen for the
MLT parameter $\alpha=l/H_{p}$ (where $H_{p}$ is the pressure scale height
and $l$ the mixing length). All atomic and molecular level populations are
computed under the assumption of LTE.

Line--blanketing is introduced using the Opacity Sampling formalism
\citep{pet74,sn76}. Opacities come from various sources (Kurucz, TOPbase,
VALD) or were computed by the MARCS team. They are listed in Sect.\ 4 of
\citet{gus08}. In addition, molecular line lists are constantly added in
order to improve the quality of the models and to solve problems related to
specific stars and spectroscopic features.

The final synthetic spectra were computed with the code TURBOSPECTRUM. It
is an enhanced version of the ``Spectrum'' package developed at Uppsala
observatory and is widely based on the MARCS routines
\citep{plez93,ap98}. TURBOSPECTRUM uses MARCS models as input from which it
computes a formal solution of the radiative transfer
equation. Microturbulence is introduced to correctly reproduce line
profiles, and is the same as that used in the associated MARCS model
atmosphere, although this consistency is not critical (see in Gustafsson et al. 2008 the effect of microturbulence 
on the thermal structure of the models).

\subsection{ATLAS12 and SYNSPEC}
\label{s_atlas}

In the range of 7\,000 K$\le$ \teff\ $\le$ 15\,000 K, we have used ATLAS12
model atmospheres \citep{kurucz05a}. This new version of the ATLAS LTE
blanketed model atmospheres handles line opacity in stellar atmosphere
using the Opacity Sampling technique. The models assume a plane
parallel geometry, hydrostatic and radiative equilibrium, and LTE. Models
from ATLAS12 can be generated for whichever individual abundances and
microturbulent velocity. Convection in ATLAS12 also relies on
the MLT formalism. Specifically, we have adopted Smalley's
prescriptions \citep{sma04} for the values of the ratio of the mixing
length to the pressure scale height ($\alpha$, see above). We have fixed
the microturbulent velocity (constant with depth) to 2 \kms\ for all the
models.

Continuous opacities are the same as in ATLAS9 and they are mainly: HI,
H$^{-}$, H$_{2}^{+}$, He I, He II, He$^{-}$ and a number of metallic
absorbers such as C I-IV, Mg I-II, Al I, Si I-II,Fe I, Ca II, N II-V, O
II-VI, Ne I-VI.  Electron scattering, H-Rayleigh scattering, H$_{2}$
Rayleigh scattering and He I Rayleigh scattering are also included.

The line opacity is computed starting from all the atomic and molecular
lines of line lists from Kurucz's CD-ROMs.\\
The synthetic spectra are finally computed using the SYNSPEC tool.
SYNSPEC \citep{hl92} is a general LTE spectrum synthesis program that reads
an ATLAS model atmosphere and a general line list and dynamically selects
lines which contribute to the total opacity, based on physical parameters
of the actual model atmosphere. SYNSPEC then solves the radiative transfer
equation, wavelength by wavelength, in a specified wavelength range, and
with a specified wavelength resolution. The microturbulence velocity was
fixed at 2 \kms.

For the computation of a HRSS, the line list used in SYNSPEC is the same as the one used in TURBOSPECTRUM for cool stars. 
On the other hand, in case of the SEDs, we used Kurucz's gfall.dat\footnote{http://kurucz.harvard.edu/LINELISTS/GFALL/} list, 
from which we selected the lines between 9\,00 and 20\,000 \AA.

\subsection{CMFGEN and CMF\_FLUX}
\label{s_cmfgen}

Synthetic spectra of stars hotter than 25\,000 K are computed using
the codes CMFGEN and CMF\_FLUX. CMFGEN is developped by John Hillier at the
University of Pittsburgh. An exhaustive description can be found in
\citet{hm98}. A web page
(http://kookaburra.phyast.pitt.edu/hillier/web/CMFGEN.htm) provides further
information and regular updates.

CMFGEN allows the computation of non--LTE, line--blanketed atmospheres
models including stellar winds. It solves, in 1D, the statistical and
radiative transfer equations assuming spherical geometry. A co--moving
frame is used. CMFGEN is not a hydrodynamic code: the density structure
must be specified as an input. In practice, the pseudo--photospheric
structure resulting from TLUSTY models \citep{tlusty03} is used and is
connected to a so--called $\beta$--velocity law \citep[e.g.][]{ll99} to
give the input velocity structure. The density structure directly follows
from mass conservation. The $\beta$--velocity law has been shown to be a
good approximation of the outer structure of the winds of hot massive stars
\citep{cak75,pauldrach86}. Sixty depth points are usually included for a
good sampling of the input structure. Recently, a partial iteration of the
hydrodynamic structure in the inner atmosphere has been implemented
\citep[see a brief description in][]{wr09}. This is especially relevant for
Wolf-Rayet stars for which SED and SSHR data will be soon included in the
database.

 CMFGEN allows the inclusion of clumping with the following formalism.  A
 volume filling factor $f$ is assumed to vary monotonically with depth,
 starting from a value of 1 at the photosphere and reaching a maximum value
 $f_{\infty}$ at the outer boundary of the atmosphere. In practice, $f$ is
 parametrized as a function of the local wind velocity $v$

\begin{center}
\begin{equation}
f = f_{\infty} + (1-f_{\infty})e^{-\frac{v}{v_{cl}}}
\label{eq_clump}
\end{equation}  
\end{center}

\noindent where $v_{cl}$ is a parameter fixing the velocity at which
clumping starts to become significant. Values for $v_{cl}$ vary between 30
and 200 \kms. The filling factor given in the POLLUX header is $f_{\infty}$
and is classically close to 0.1 \citep[][]{hil91,hil03,jc05,wr09}.

Line--blanketing is included using the super--level approximation: levels of similar energies are grouped into a single super--level 
which is used to compute the statistical and radiative transfer equations. Within a super--level, the populations of individual 
levels are assumed to follow LTE conditions. Models without any super--level (i.e. all levels are treated individually) can in 
principle be computed, but the amount of RAM necessary is prohibitive. In practice, standard models currently include: 
H, He, C, N, O, Si, S, Fe. Other elements such as Ar, Ca, Ne, Ni are sometimes included for a better treatment of line--blanketing. 
However, line--blanketing effects are mainly due to Fe \citep[e.g.][]{msh02}. Typical models include 2\,000 to 5\,000 levels 
grouped in about 1\,000 to 2\,000 super--levels, requiring 1 to 3 Gb of RAM.  

For the computation of the atmospheric structure, Doppler line profiles are
used. Broadening by a microturbulent velocity can be added. This velocity
is fixed within the atmosphere. Values of 5 to 50 are chosen depending on
the type of star. Atomic data are mainly taken from the Opacity Project
database \citep{badnell05,seaton05} and are occasionally complemented by
specific line lists.

Once obtained, the atmosphere structure is held fixed and a formal solution
of the radiative transfer equation in the observer's frame is performed to
yield the final synthetic spectrum. The code CMF\_FLUX is used
\citep{hm98}. Voigt line profile including Stark broadening are specified
for individual lines. A depth--variable microturbulent velocity can be
used, starting from a minimum value at the photosphere (5 to 50 \kms) and
increasing linearly to a maximum value (usually 10\% of the maximum wind
velocity). The detailed spectrum and the theoretical continuum can be
computed separately, allowing the production of normalized spectra. In the
case of CMFGEN/CMF\_FLUX models, both the HRSS and the SED data result from
CMF\_FLUX computations.

\subsection{Science with POLLUX data}

The spectra (and SED) computed with the codes described above allow at
present a partial coverage of the HR diagram as can be seen on
Fig.~\ref{teff_logg_FeH0}. The database does not include spectra nor SED
for early B type stars. This gap should partly be filled in forthcoming
releases for B-type supergiants with spectra based on CMFGEN models. On the
other hand, main sequence B-type stars fall at the lower (upper)
temperature limit of the validity of CMFGEN (ATLAS) models, and neither
codes are really adapted to provide spectra for these objects. Concerning
the transition between G and F-type stars, we do not include any overlap
between the warm models based on ATLAS atmospheres and the cool models
based on MARCS atmospheres. The lack of overlap in the present version of
the database denotes the limits of full validity of the model atmospheres
that have been used so far to generate the spectra in POLLUX. Actually, a
series of works based on ATLAS9 model atmospheres have been published in
the last decade providing SEDs and synthetic spectra down to low
temperatures \citep{Munari99,Munari00,Castelli01,Bertone2008} with
reasonable success. However, the use of ATLAS models to analyze stars with
temperatures lower than 7000 K is hampered by the incomplete treatment of
molecules, as parlty discussed by \cite{Kurucz05b}. This is the main reason
why in the POLLUX database, the ATLAS code has been used only for effective
temperature larger than 7000 K, ensuring that molecules can be
neglected. \\ The data (SEDs and HRSS) available in the POLLUX database
have been designed to serve several purposes, from population synthesis to
fundamental parameters determinations. The physics used in the model
atmospheres and the spectral synthesis codes have been validated by direct
comparison to observed spectra in each range of temperature:
 \begin{itemize}
\item CMFGEN/CMF\_FLUX models have been repeatedly and successfully
compared to observed spectra to derive wind properties and stellar
parameters for O-type and Wolf-Rayet stars
\citep{Hillier2003,Bouret2003,Martins2005,Bouret2008,wr09}. Part
of the CMFGEN SEDs included in the POLLUX database have been used by \cite{MartinsPlez06}
to build the most recent calibration of the
photometry of O stars. Two systems have been adopted: Johnson-Cousins for
the optical range (UBV) and Johnson-Glass for the near-infrared (JHK). The
resulting calibrations provides a set of magnitudes and colors as well as
bolometric corrections for a given spectral type. These calibrations are
widely used in massive stars studies to infer luminosities of newborn,
embedded massive stars and bona-fide O stars \citep[e.g.][]{Zavagno07,Deharveng09}.
\item The version ATLAS9 of the ATLAS/Kurucz code for model atmospheres has
  been compared to observed spectra and combined with the SYNSPEC or the
  SYNTHE spectral synthesis modules to determine fundamental parameters and
  chemical abundances
  \citep{Castelli1997,Castelli1999,Gebran08a,Gebran08b}. It has also been
  used to compute synthetic colors successfully confronted to the colors of
  real stars \citep{Castelli1999}. On the other hand, there are only a few
  works published using the newer ATLAS12 version of ATLAS, where major
  modification of the treatment of opacities have been implemented. We may
  however cite \cite{Castelli2009,Hubrig2009}, who have used ATLAS12/SYNTHE
  for abundance and stellar parameters determinations in chemically
  peculiar B-type stars. As shown by \cite{Gebran07} and
  \cite{Gebran08a}, the atmospheric structure of plan-parallel LTE models
  computed for A and F type stars with ATLAS9 and ATLAS12 codes are in very
  good agreement.
\item MARCS model atmospheres and synthetic spectra computed with TURBOSPECTRUM have been widely used 
to derive fundamental parameters and abundances of late-type stars (in
particular G and later types) . 
The comparison to observed spectra has shown the ability of these models to reproduce both SEDs and high-resolution 
spectra  \citep{ap98,Edvardsson08,Plez08}. 
Synthetic photometry has been calculated for stars ranging from dwarfs to
giants by \cite{Bessell98} and \cite{Buzzoni2010} for Johnson-Cousins passbands, or by \cite{Onehag09} for Str\"omgren filters, that 
successfully match observations. Detailed spectrophotometric fits could be
obtained for red supergiants in the Galaxy, the Magellanic Clouds, and M31
by \cite{Levesque05,Levesque06,Levesque07} and by \cite{Massey08,Massey09}. Red supergiants
parameters derived from these calculations lead to evolutionary sequences matching observations.
A very large number of contributions to studies of nucleosynthesis and chemical evolution have used MARCS model spectra. 
They include the  derivation of abundances for large samples of Galactic stars \citep[e.g.][]{Edvardsson93}, of very metal-poor stars 
\citep[e.g.][ and references therein]{Bonifacio09}, and of 
carbon stars \citep[e.g.][]{Wahlin06} and AGB stars \citep{GH07}. 
\end{itemize}


\section{The POLLUX web site}
\label{s_web}

The POLLUX database is accessible via the URL: 
\underline{\bf http://pollux.graal.univ-montp2.fr}. Below, we describe how to use the web interface to browse through 
the database and retrieve data.

The web page and the database have been elaborated using Plone and Python
languages. The web page is designed to be user friendly, and to allow
inexperienced users to easily access, visualise and retrieve the data. 
The homepage contains a documented user's guide and a link labelled \underline{\bf Access to the POLLUX Database} that leads to the POLLUX query form.

\subsection{Query Interface}

\noindent The POLLUX query form is divided into two parts. On the left side, a tree selection area allows to define the 
general database parameters,  by choosing :
\begin{itemize} 
\item[--] the {\sf Type of Data} (high resolution spectrum, SED or both)
\item[--] the {\sf Model Atmosphere} the data are based on (MARCS, CMFGEN,
  ATLAS or all) 
\item[--] the {\sf Type of Model Atmosphere} (plane-parallel, spherical or both).
\end{itemize}
This first query is hierarchical. The default query concerns the entire database (SED+HRSS from 
MARCS, ATLAS12  and CMFGEN, for 
both  plane-parallel and spherical model atmospheres).
 
Once the choice of the general database parameters has been made, the user must specify the spectrum parameters.
The allowed query depends on the Model Atmosphere Family that
has been chosen from the tree query section, and then some irrelevant fields may automatically appear as frozen. 
At that stage, the different spectrum parameters that can be selected are :
\begin{itemize} 
\item[--] effective temperature 
\item[--] log g
\item[--] mass (irrelevant for data derived from MARCS and ATLAS
  plane-parallel model atmospheres)
\item[--] luminosity (irrelevant for data derived from MARCS and ATLAS
  plane-parallel model atmospheres)
\item[--] microturbulent velocity
\item[--] metallicity [Fe/H]
\end{itemize}

\noindent The user must either choose a value interval or choose an exact
value, for {\em at least one} of these spectrum parameters.  If the exact
value requested is not available in the database, an error message will be
returned. The lowest and highest authorized values are indicated along the
query zones.

A third optional query block is also available, which enables the
user to choose data sets with specific abundances such as Carbon, Nitrogen, Oxygen, $\alpha$-, $r$- and
$s$- elements.

A final block indicates the Cart Status (the designed facility to download files, see Section~4.3). 
The possibility to clear cart is proposed at that stage. 

\subsection{Result of Request}

The result of the request consists of a table (dispatched on
several pages if needed) containing three main parts :  Display , 
  Data Characteristics and  Cart.

\subsubsection{Display}

Four columns are ordered as follows:
\begin{enumerate}
\item Selection zone allowing (through the OVERPLOT command) the graphical display of one to three spectra in a popup window,
\item Clickable icon allowing the graphical display of the  spectrum
  in a popup window, 
\item Clickable icon allowing the graphical display of the  spectrum
  normalised to the continuum {\em if relevant} in a popup window, 
\item Clickable icon allowing the display of the  ASCII header file
in a popup window.
\end{enumerate}

\noindent The graphical tool that allows the visualisation of the SED and
spectrum integrates a pointer and different zoom
facilities. It also includes a help button. 
Fig.\ \ref{fig_web_plot} shows a snaphshot of a representative window.The image displayed in
 this window can be saved in a png format. The overplot tool displays
 simultaneously one to three spectra, in absolute flux or flux relative to
 continuum. It offers the same functionality (pointer, zoom, ...) as the
 basic graphical tool.

\begin{figure*}[th]
\label{fig_web_plot}
\centering
\includegraphics[width=10cm]{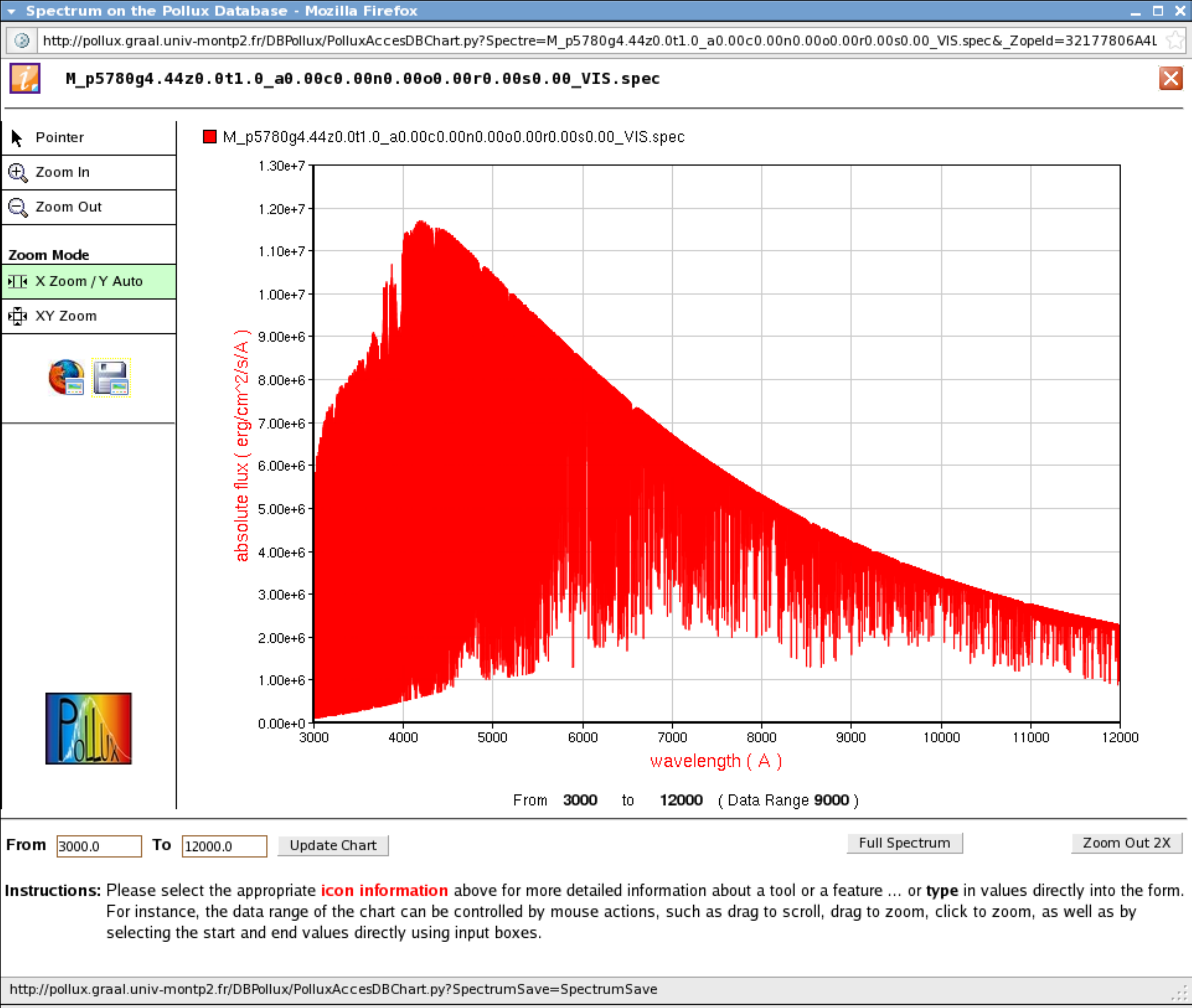}
\caption{Graphical tool for spectrum visualisation. Different zoom options are available. }\label{fig_web_plot}
\end{figure*}

\subsubsection{Data Characteristics}

Columns 5 to 19 of the table present the data characteristics, 

\begin{itemize}
\item[5.] {\bf Data Type} (of model atmosphere)
\item[6.] {\bf Model} (atmosphere family)
\item[7.] {\bf Type} (of model atmosphere)
\item[8.] {\bf  T$_{eff}$}
\item[9.] {\bf log g}
\item[10.] {\bf M / M$_\odot$}
\item[11.] {\bf log L/L$_\odot$}
\item[12.] {\bf microturbulent velocity}
\item[13.] {\bf [Fe/H]}
\item[14.] {\bf [$\alpha$-elements/H]}
\item[15.] {\bf [C/H]}
\item[16.] {\bf [O/H]}
\item[17.] {\bf [N/H]}
\item[18.] {\bf [$r$-elements/H]}
\item[19.] {\bf [$s$-elements/H]}
\end{itemize}

\subsubsection{Cart}

The last two columns of the table allow 
\begin{itemize}
\item[20.] either to directly download a {\bf
  gzipped tar archive {\em filename.tgz} } containing the pair
  spectrum (or SED) + associated header in ASCII format  
\item[21.] or to select various rows and fill a cart with the data that
  the user wishes to download.
\end{itemize}

\noindent The column with the cart ends with a cart icon that, when
clicked, leads to the retrieval page.

\subsection{Cart and Retrieval}

\noindent This last page is divided into two parts.
In its central section, a table lists the data stored in the cart.
 The columns are ordered as follows :
 
\begin{enumerate}
\item Select or unselect the file which will actually be included in the
  downloadable archive.
\item Icon to remove the data set from the cart.
\item Filename (as explicitly described in Fig.~1).
\item Type of data (Spectral Energy Distribution or High Resolution Synthetic Spectrum).
\item Size of the compressed archive file that will be retrieved (this file
  contains the data set and the associated ASCII header file). 
\end{enumerate}

\noindent On the left side of the page, the user can choose the file
and archive type. As of December 2009, the available formats for the
data are  Flat Table (ASCII),  XML VOTable (including binary XML VOTable), and  FITS. 
The archive can either be a {\it tar} file or a
{\it zip} file. The total size of the archive is also given.
In the  Flat Table format (ASCII files), the user will retrieve
a pair of files for each data set, namely the actual data file and its associated header, the
name of which is the same except for the final extension.


\section{Orientation towards the Virtual Observatory}
\label{s_VO}

\begin{figure*}[ht]
\centering
\includegraphics[width=10cm]{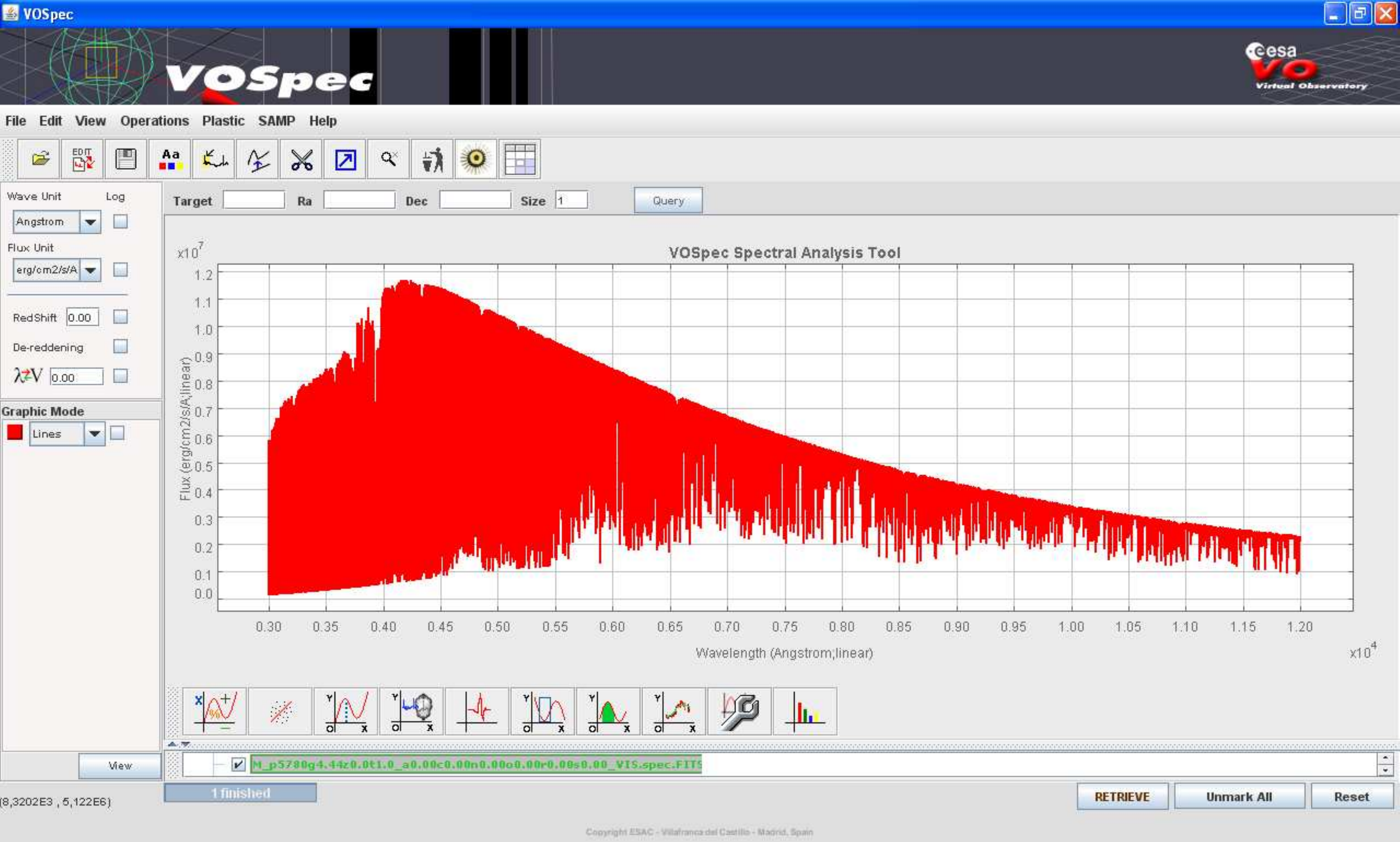} 
\caption{Graphic display of the POLLUX MARCS solar spectrum (HRSS MARCS
  data) through VOSpec tool {\bf using} the Theoretical Spectra Access Protocal (TSAP).}
\label{VOspec-sun}
\end{figure*}
 
\noindent The evolution of POLLUX towards a Virtual Observatory (VO) compliant service has recently started and is still 
under promising development. The POLLUX data can now be retrieved in various formats, including VOTable and FITS 
compatible with the standards of the VO. Files in format XML VOTable, XML Binary VOTable and FITS have been generated 
using the TOPCAT tool (Tool for OPerations on Catalogues And Tables\footnote{http://www.star.bristol.ac.uk/~mbt/topcat/}). 

\noindent The POLLUX spectra can already be handled through some services
of the Virtual Observatory, such as
VOSpec\footnote{http://esavo.esa.int/vospec/} (see Fig.~\ref{VOspec-sun})
or Aladin\footnote{http://aladin.u-strasbg.fr/aladin.gml}. POLLUX is
registered in the different registries\footnote{In particular in the
EURO-VO registry (http://registry.euro-vo.org/) and the NVO registry
(http://nvo.stsci.edu/vor10/index.aspx).} 
as a service using the Simple
Spectra Access Protocol (SSAP, \citealt{tody08}) and the Theoretical
Spectra Access Protocol (TSAP). The SSAP defines a uniform interface to
track and access 1-D spectra. However, being originally designed to access
observational data, it cannot be used directly to query the POLLUX
database. A way to work around the impossibility to locate (with
coordinates on the sky) the synthetic spectra is to use the FORMAT=METADATA
mechanism to query theoretical data in the context of the SSAP. This
corresponds to the TSAP, that has been included in the SSAP as a use
case\footnote{see Appendix A in http://www.ivoa.net/Documents/
latest/SSA.html}, and that allows to make a query and visualize data from
the POLLUX database using the VOSpec tool as shown in
Fig.~\ref{VOspec-sun} for the POLLUX MARCS solar spectrum in
units of absolute flux versus wavelength. 

All the relevant characteristics of the data, appearing in the POLLUX query forms, 
as well as those listed in the header files coming with any POLLUX spectra, have been associated to an Unified Content 
Descriptor
\citep[UCD\footnote{http://www.ivoa.net/Documents/latest/UCD.html}, ][]{derr05} that 
ensures the data interoperability within the VO context.

\noindent Finally, the Pollux database has been registered as a VO service providing theoretical 
spectra that can be handled, through VO tools, 
within the framework of Science Cases. In particular two scientific applications are already in progress :

\begin{enumerate}

\item The connexion to spectroscopic or spectropolarimetric instruments archives, such as the Castor 
  project\footnote{http://magics.bagn.obs-mip.fr/indexCastor.html},
  the legacy database of spectropolarimetric observations collected
  with the ESPaDOnS and NARVAL instruments.  The link between these
  two databases (providing respectively theoretical and observed
  ressources) will be insured with the help of the MATISSE algorithm
  \citep{recio06}.  Using a selection of theoretical POLLUX spectra as
  reference, this automated algorithm will derive stellar atmospheric
  parameters and chemical abundances (\teff, \logg , global
  metallicity $[M/H]$ and $[\alpha/Fe]$) for each Castor spectrum.

\item An application with the Meudon PDR
  code\footnote{http://pdr.obspm.fr/PDRcode.html} \citep{LePetit06}.
  This code will compute the atomic and molecular structure of
  interstellar clouds, using a theoretical POLLUX SED to light up the
  stellar formation region. In a second step, a coupling of the PDR
  code to theoretical POLLUX spectra and to observed FUSE data is
  foreseen. Through several VO tasks, the aim is to simulate a
  complete line of sight, from the stellar luminous source to the
  absorbing diffuse medium, and to compare it to available FUSE
  observations.

\end{enumerate}

\section{Upcoming developments}
\label{s_future}

\noindent At present non-solar metallicity data are only available in
POLLUX for late B, A and F spectral types (7000 $K \le$ \teff~$\le 15000 K$). Data at sub-solar metallicities for cool 
stars (G to M types) should be made available in a forthcoming
release. A large dataset of HRSS based on the latest generation of
  MARCS model atmospheres produced for scientific purposes increasing the
  HR diagram coverage could be included shortly.
In particular, new SED and HRSS data sets derived from the MARCS model atmospheres will be added at 
[Fe/H] = -1.0, -2.0, -3.0, -4.0 and -5.0. Additional data with non-solar CNO abundances will also be made available
for cool stars. Data with enhanced [C/Fe] are already available in the
effective temperature range \teff $\in$ [4000 $K$; 6000 $K$].

\noindent Concerning hot and massive stars, additional data (HRSS and SED)
related to B-type supergiants and Wolf-Rayet stars will be derived from CMFGEN model
atmospheres and progressively included in the database. The introduction of
Wolf-Rayet spectra requires a modification of the query interface and the
creation of specific header files due to the importance of parameters that
describe exclusively these kind of objects. This modification has already
been done anticipating this development.

\noindent In a second time, high resolution synthetic spectra covering the infrared domain will be progressively made available. 
However, this step still requires improvements in the production of atomic and molecular line lists.

\begin{acknowledgements}
We warmly thank D.J. Hillier for making his code CMFGEN available to the
community.
The POLLUX team warmly thanks P. Maeght for his help in the process of
including POLLUX in the Virtual Observatory.
We acknowledge financial support from "Programme National de Physique Stellaire" (PNPS) and "Action Sp\'ecifique Observatoire Virtuel" (ASOV) of CNRS/INSU, France. 
\end{acknowledgements}

\bibliography{13932_main}

\end{document}